\documentclass[aps,twocolumn]{revtex4}
\usepackage[utf8]{inputenc}
\usepackage{amsmath}
\usepackage{amsfonts}
\usepackage{amssymb}
\usepackage{graphicx}
\usepackage{epstopdf}

\begin{document}

\title{Correlation transfer in\textbf{\ }large-spin chains}
\author{I.F. Valtierra $^1$, J.L. Romero $^1$, and A.B. Klimov$^{1,2}$}
\affiliation{$^1$Departamento de F\'isica, Universidad de Guadalajara, Revoluci\'on 1500,
Guadalajara, Jalisco, 44420, M\'exico.\\
$^{2}$Center of Quantum Optics and Quantum Information, Center for Optics
and Photonics, Departamento de F\'{\i}sica, Universidad de Concepci\'{o}n,
Casilla-160C, Concepci\'{o}n, Chile.}

\begin{abstract}
It is shown that transient spin-spin correlations in one-dimensional spin $%
S\gg 1$ chain can be enhanced for initially factorized and individually
squeezed spin states. Such correlation transfer form "internal" to
''external'' degrees of freedom can be well described by using a semiclassical
phase-space approach.
\end{abstract}
\maketitle

\section{ Introduction}

Spin chains provide a natural mechanism for creation of quantum channels
targeted on correlating of spatially separated particles. State \cite%
{PhysRevLett.70.189,bose2007quantu,PhysRevLett.91.20790,horo,baya} and
entanglement \cite%
{friedan1986conforma,subrahmanyam2006transpor,plenio2004dynamic,PhysRevA.75.06232,PhysRevA.76.05232,amico,osborne2002entangleme}
transfer can be achieved even in the simplest types of spin chains, which
allows to employ them as a medium for a short distance quantum
communication. Special attention attracts the idea (recently realized
experimentally \cite{ex}) of entanglement generation between distant spins 
\cite{venut,Pop,yung,wich,franco,wang,zueco}. It was noted \cite%
{bayat2007transfe} that while the quality of entanglement transfer in spin
chains increases with the dimension of interacting spins, the average
fidelity still significantly diminishes with the distance between the spins.
In order to enhance the pairewise entanglement we propose to use
correlations stored in the initial \textit{factorized} state of a chain
consisting of large size spins. The main idea is to prepare each spin of the
chain in an appropriate squeezed state, so that the correlation from the
"internal degree of freedom" is transferred into the spin-spin entanglement
during the evolution of the system. It worth noting, that the possibility of
correlation transfer between subsystems of various physical systems has been
widely discussed both from theoretical \cite{C,cubit,retama,wan} and
experimental \cite{exp C,Dav,le,vaz,barr} perspectives.

This article is organized as follows: first we show that for a particular
type of spin-spin interaction it is possible to choose an optimal initial
spin squeezed state, so that the maximum amount of entanglement between
neighbors spins, as well as the correlation of a spin with the rest of the
particles in the chain substantially increase; then, we argue that the
dynamics of the correlation transfer for large spins can be fairly well
described from the semiclassical point of view using the language of the
phase-space distributions, which opens the possibility for large-scale
simulations of correlation dynamics in spin chains.

\section{The model}

Let us consider an open chain of $N$ spins of size $S$ with homogeneous
Ising-like interaction (the coupling constant is taken to be unity),
governed by the Hamiltonian 
\begin{equation}
H=\sum_{j=1}^{N}S_{z_{j}}S_{z_{j+1}}.  \label{hami}
\end{equation}%
Initially the spins are prepared in a factorized state%
\begin{equation*}
|\psi _{0}\rangle =\prod\limits_{j=1}^{N}|\psi _{0}\rangle _{j},
\end{equation*}%
where each spin is in the coherent state localized on the equator of the
Bloch sphere subjected to a squeezing transformation (generated by $%
S_{z}^{2} $) and a rotation around $x$-axes in the maximum squeezing
direction (in the corresponding tangent plane), 
\begin{eqnarray}
|\psi _{0}\rangle _{j} &=&e^{i\vartheta S_{x_{j}}}e^{-i\mu
S_{z_{j}}^{2}}|\pi /2,0\rangle _{j},  \label{psi0} \\
|\pi /2,0\rangle &=&\frac{1}{2^{s}}\sum\limits_{k=-S}^{S}\sqrt{\frac{2S!}{%
\left( S-k\right) !\left( S+k\right) !}}|k,S\rangle .  \notag
\end{eqnarray}

Since $|\psi _{0}\rangle $ is not an eigenstate of the Hamiltonian (\ref%
{hami}) some specific spin-spin correlations arise during the Hamiltonian
evolution. The state of the system at time $t$ is

\begin{equation}
|\psi \left( t\right) \rangle =\prod_{j=1}^{N}\sum\limits_{k=-S}^{S}\gamma
_{k_{j}}E_{k_{j}}\left( t\right) P_{m,k}^{(j)}\left( \vartheta \right)
|m,S\rangle _{j},  \label{theta}
\end{equation}%
where 
\begin{eqnarray*}
P_{m,k}^{(j)}\left( \vartheta \right) &=&\,_{j}\langle m,S|e^{i\vartheta
S_{x}}|k,S\rangle _{j},\quad \gamma _{k}=\frac{1}{2^{S}}\sqrt{\frac{(2S)!}{%
(S-k)!(S+k)!}}, \\
E(t)_{k_{j}} &=&\exp \left( -i t\sum_{l}k_{l}m_{l+1}\right)
e^{-i\mu k_{j}^{2}}.
\end{eqnarray*}

The simplest characteristics of a correlation between a selected spin and
the rest of the spin in the chain is the\textbf{\ }\textit{I}-Concurrence 
\cite{PhysRevLett.80.224}, 
\begin{equation}
\mathcal{C}_{I_{j}}=\frac{2S+1}{2S}(1-\mathcal{P}_{j}\left( t\right) ),
\label{concurrence}
\end{equation}%
where $\mathcal{P}_{j}(t)=\mathrm{Tr}(\rho _{j}^{2}(t))$ is the purity of
the $j$-th spin. The degree of entanglement between a pair of spins can be
described by the negativity \cite{PhysRevA.65.032314}, 
\begin{equation}
\mathcal{N}(\rho _{kl})=\frac{2S+1}{2S}(||\rho _{kl}^{tk}||_{1}-1),
\label{negativity}
\end{equation}%
where $||\rho ||_{1}=Tr\sqrt{\rho \rho ^{\dag }}$ is the trace norm of the
partially transposed (on the $k$-th spin) reduced two-particle ($k$-$l$)
density matrix $\rho _{kl}^{tk}$.

\section{Optimization of the correlation transfer}

We have numerically optimized the $\mathit{I}$-Concurrence with respect to
the squeezing parameter $\mu $ and the rotation angle $\vartheta $ for a) a $%
j$-th spin, $j\neq 1,N$; b) the last (first) spin in the chain.

\begin{figure}[tbp]
\centering
\includegraphics[width=7cm]{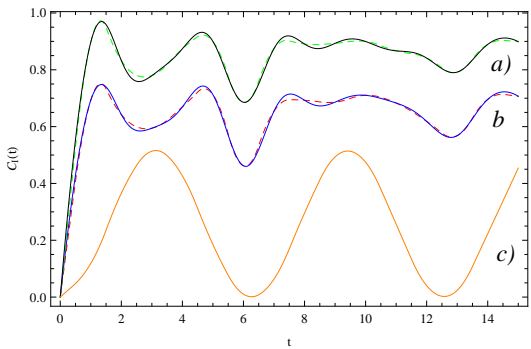}
\caption{The evolution of the $C_{I_{j}}$ in the chain of 6 spin of size $%
S=5 $. The upper curve corresponds to the optimum squeezing case with $%
\protect\mu _{opt}=0.5$ and $\protect\vartheta =0.2645$. The middle curve
represents non-squeezed initial spin coherent state. The lowest curve is $%
C_{I_{j}}$ for $S=1/2$. Exact calculations are represented by solid lines
and the semi-classical approximation by dotted lines. }
\label{fig:mil31}
\end{figure}

\begin{figure}[tbp]
\centering
\includegraphics[width=7cm]{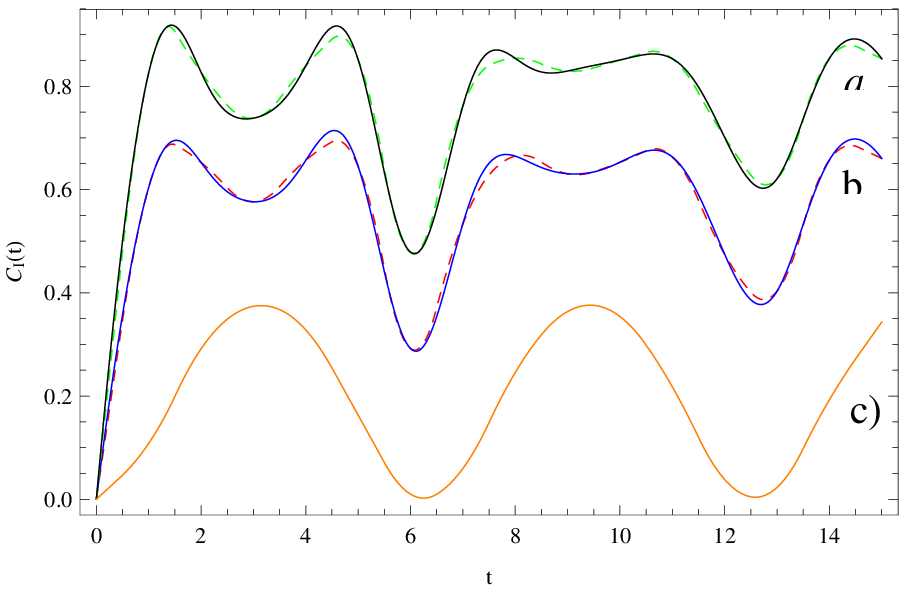}
\caption{The evolution of the $C_{I_{N}}$ in the chain of 6 spin of size $%
S=5 $. The upper curve correpsonds to the optimum squeezing case with $%
\protect\mu _{opt}=0.5$ and $\protect\vartheta =0.2643$. The middle curve
represents non-squeezed initial spin coherent state. The lowest curve is $%
C_{I_{N}}$ for $S=1/2$. Exact calculations are represented by solid lines
and the semi-classical approximation by dotted lines.}
\label{fig:mil32}
\end{figure}

It results that the optimum squeezing parameter is \textit{the same} in both
cases and scales as $\mu _{opt}\simeq 0.126\ S^{-0.858}$; the optimum
rotation angle behaves as $\vartheta _{opt}\simeq \frac{1}{2}\mathrm{arctan}%
\ {S^{-1/3}}$. For $j$-th spin, $j\neq 1,N$, $\max C_{I_{j}}$ scales with
the number of spins in the chain as $\mathcal{\sim }N^{1/5}$ and depends on
the size of spins as $\max C_{I_{j}}\simeq A^{g(S)}$, where $A=0.681$ and $%
g(S)=S^{1/5}$. For the last spins in the chain $\max \mathcal{C}_{I_{1,N}}$
scales with the number of spins as $\sim N^{1/10}$ and depends on the spin
size as $\max C_{I_{1,N}}\simeq B^{f(S)}$ where $B=0.523$ and $%
f(S)=S^{0.191} $. It is worth nothing that the instants, when the maximum
value in the $I$-concurrence is reached does not depend nor on the spin size 
$S$, neither on the longitude of the chain. The dynamics of the $I$%
-concurrence for the optimized initial spin squeezed (upper curve) and
non-squeezed, $\mu =0$, (middle curve) states in case an "internal" spin is
plotted in Fig. (\ref{fig:mil31}) and for the last spins of the chain in (%
\ref{fig:mil32}), here $S=5$, $N=6$ .

The lowest curve represent the $I$-concurrence for spins-$1/2$. One may
observe that a certain improvement of the entanglement between spins is
achieved by inducing correlations in the initial \textit{factorized} state
of the system. This effect becomes even more pronounced for internal spins
in the limit of large spins $S\gg 1$, when $C_{I_{j}}$ tends to a constant
value close to unity.

\begin{figure}[tbp]
\centering
\includegraphics[width=7cm]{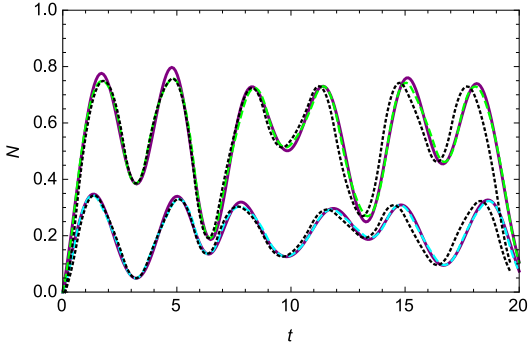}
\caption{The evolution of $N_{j,j+1}$ for an internal pair, $N=7$, $S=7$.
The upper curve correpsonds to the optimum squeezing case with $\protect\mu %
_{opt}\approx 0.856S^{-0.72}$. The lowest curve represent the evolution
without squeezing. Exact calculations are represented by solid lines; the
semi-classical approximation using (\protect\ref{N}) and (\protect\ref{NN})
by dotted and dashed  lines correspondingly.}
\label{fig:nega1}
\end{figure}

\begin{figure}[tbp]
\centering
\includegraphics[width=7cm]{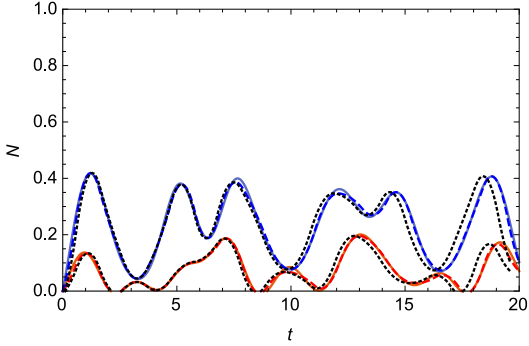}
\caption{The evolution of $N_{1,2}$, $N=7$, $S=7$. The upper curve
correpsonds to the optimum squeezing case with $\protect\mu _{opt}\approx
0.856S^{-0.72}$. The lowest curve represent the evolution without
squeezing.Exact calculations are represented by solid lines; the
semi-classical approximation using (\protect\ref{N}) and (\protect\ref{NN})
by dotted and dashed lines correspondingly.}
\label{fig:nega2}
\end{figure}

In the one-fimensional chain (\ref{hami}) the negativity (\ref{negativity})
takes non-zero value only for consequtive spins. The negativity $\mathcal{N}%
_{j,j+1}$ is optimazed for the squeezing parameter $\mu _{opt}\approx
0.856S^{-0.72}$, but scales in a different way for a1) internal pairs and
for 2) external pairs, i.e. when one of the spins is located at the end of
the chain: in the first case $N_{j,j+1}\sim N^{-0.13}$\ and \ $N_{j,j+1}\sim
D^{h(S)}$,\ where\ $D=0.721$\ and\ $h(S)=S^{0.047}$; in the second case $%
N_{1,2}\sim N^{-0.45}$\ and $N_{1,2}\sim G^{r(S)}$, where \ $G=0.341$\ and\ $%
h(S)=S^{0.1}$. The evolution of $N_{j,j+1}$ for the optimized initial spin
squeezed (upper curve) and non-squeezed, $\mu =0$, (middle curve) states in
case an internal pairs is plotted in Fig. (\ref{fig:nega1}) and for the
externail pairs (\ref{fig:nega2}), here $S=7$, $N=7$.

\textbf{\ } 

\section{Semiclassical approach}

Unfortunately, the numerical optimization becomes quite involved for
non-diagonal spin chain Hamiltonians and large spin size. In order to be
able to carry out such optimization it is appropriate to use the phase-space
methods. According to this approach we can reformulate the quantum mechanics
on the language of distributions (symbols of operators) in the classical
phase-space, where both states and observables are considered as smooth
functions in such a way that average values are computed by convoluting
symbols of the density matrix and the corresponding operators \cite{Wi}.

In addition, the quantum dynamics can be efficiently simulated in terms of
the spin Wigner function defined as an invertible map\ of the density matrix 
$\rho $\ to a smooth distribution $W_{\rho }(\Omega )$ on the
two-dimensional sphere $\Omega =(\theta ,\phi )\in S_{2}$

\begin{eqnarray}
W_{\rho }(\Omega ) &=&\mathrm{Tr}\left( \rho \,\hat{\omega}(\Omega )\right) ,
\label{Wex} \\
{\rho } &{=}&\frac{2S+1}{4\pi }\int_{\mathcal{S}_{2}}d\Omega \,\hat{\omega}%
(\Omega )W(\Omega )
\end{eqnarray}%
where $\hat{\omega}(\Omega )$ is the kernel operator \cite{wigne}, 
\begin{equation}
\hat{\omega}(\Omega )=\frac{2\sqrt{\pi }}{\sqrt{2S+1}}\sum_{L=0}^{2S}%
\sum_{M=-L}^{L}Y_{LM}^{\ast }(\Omega )\hat{T}_{LM}^{(S)},  \label{eq:kernel}
\end{equation}%
being $Y_{LM}(\Omega )$ the spherical harmonics and $\hat{T}_{LM}^{(S)}$ the
irreducible tensor operators \cite{vars}.\newline

In the large spin (semiclassical) limit, $S\gg 1$, the Wigner function
satisfies the Louville equation \cite{klimo}, 
\begin{equation}
\partial _{t}\;W_{\rho }(\Omega )=2\epsilon \{W_{\rho }(\Omega
),W_{H}(\Omega )\}+O(\epsilon ^{3}),  \label{eq:evolucion}
\end{equation}%
where $\epsilon =(2S+1)^{-1},$ and\newline
\begin{equation}
\{g,f\}=\frac{1}{\mathrm{\sin }\theta }\left( \partial _{\phi }g\partial
_{\theta }f-\partial _{\theta }f\partial _{\phi }g\right) ,
\end{equation}%
is the Poisson brackets on the sphere, being $W_{H}(\Omega )$ the symbol of
the Hamiltonian. The solution of Eq. (\ref{eq:evolucion}) is just the
classical evolution of the initial distribution, 
\begin{equation}
W_{\rho }(\Omega |t)=W_{\rho }(\Omega (-t)),  \label{eq:sol}
\end{equation}%
here $\Omega (-t)=(\theta (-t),\phi (-t))$ denotes classical trajectories on
the sphere. The mean value of any observable $\hat{f}$ is computed according
to 
\begin{equation}
\left\langle \hat{f}\right\rangle =\int_{\mathcal{S}_{2}}d\Omega
W_{f}(\Omega )W_{\rho }(\Omega ).
\end{equation}%
where $W_{f}(\Omega )$ is the Wigner symbol of $\hat{f}$.\newline
In the case of multi-partite systems the mapping kernel is a product of
individual kernels Eq.(\ref{eq:kernel}) so that,\qquad\ 
\begin{eqnarray}
W_{\rho }\left( \mathbf{\Omega }\right) &=&Tr\left( \rho \hat{\omega}%
_{1}(\Omega _{1})...\hat{\omega}_{N}(\Omega _{N})\right) ,  \label{WF2p} \\
\mathbf{\Omega } &\mathbf{=}&\mathbf{(}\Omega _{1},...,\Omega _{N}),
\end{eqnarray}%
and the Wigner function of the reduced bi-partite system of $j$ and $j\prime 
$ spins has the form%
\begin{equation}
W\left( \Omega _{j},\Omega _{j^{\prime }}\right) =\left( \frac{2S+1}{4\pi }%
\right) ^{N-2}\int d\mathbf{\Omega }^{\prime }W_{\rho }\left( \mathbf{\Omega 
}\right) ,  \label{W2}
\end{equation}%
where the integration is performed over all spins except for $j$-th and $%
j\prime $-th. Integrating $W_{\rho }\left( \Omega _{j},\Omega _{j^{\prime
}}\right) $ over one of the solid angles one obtaines a single-particle
Wigner function, e.g. 
\begin{equation}
W\left( \Omega _{j}\right) =\frac{2S+1}{4\pi }\int_{\mathcal{S}_{2}}d\Omega
_{j^{\prime }}W\left( \Omega _{j},\Omega _{j^{\prime }}\right) .  \label{W1}
\end{equation}

Taking into account that the symbol of the Hamiltonian Eq.(\ref{hami}) is

\begin{equation*}
W_{H}(\mathbf{\Omega })=S(S+1)\sum_{j=1}^{N-1}\mathrm{\cos }\theta _{j}%
\mathrm{\cos }\theta _{j+1},
\end{equation*}%
and solving the equation of motion Eq.(\ref{eq:evolucion}) one obtains the
classical trajectories in the form:\newline
$\theta _{l}=\theta _{l0}$ where $l=1,2,..N$ and\newline
\begin{eqnarray*}
\phi _{10} &=&\phi _{1}-\frac{t}{2\epsilon }\mathrm{\cos }\theta _{2},\quad
\phi _{20}=\phi _{2}-\frac{t}{2\epsilon }(\mathrm{\cos }\theta _{1}+\mathrm{%
\cos }\theta _{3}), \\
\quad \phi _{30} &=&\phi _{3}-\frac{t}{2\epsilon }(\mathrm{\cos }\theta _{2}+%
\mathrm{\cos }\theta _{4}),...,\phi _{N0}=\phi _{N}-\frac{t}{2\epsilon }%
\mathrm{\cos }\theta _{N-1}\newline
,
\end{eqnarray*}%
here $\theta _{l0}$ and $\phi _{l0}$ stand for the initial values of the
corresponding angles. Using the above results we can represent the purity
involved in Eq.(\ref{concurrence}) as follows 
\begin{equation*}
\mathcal{P}_{j}(t)=\frac{2S+1}{4\pi }\int d\Omega _{j}W^{2}(\Omega _{j},|t).
\end{equation*}%
In order to exress the negativity in terms of the Wigner functions we first
reconstruct the bi-partite density matrix from (\ref{W2}) and its partially
transposed form,%
\begin{eqnarray*}
{\rho }_{jj^{\prime }} &{=}&\left( \frac{2S+1}{4\pi }\right) ^{2}\int
d\Omega _{j}d\Omega _{j^{\prime }}W_{\rho }\left( \Omega _{j},\Omega
_{j^{\prime }}\right) \hat{\omega}(\Omega _{j})\hat{\omega}(\Omega
_{j^{\prime }}), \\
{\rho }_{jj^{\prime }}^{t_{j}} &{=}&\left( \frac{2S+1}{4\pi }\right)
^{2}\int d\Omega _{j}d\Omega _{j^{\prime }}W_{\rho }\left( \Omega
_{j},\Omega _{j^{\prime }}\right) \hat{\omega}^{T}(\Omega _{j})\hat{\omega}%
(\Omega _{j^{\prime }})
\end{eqnarray*}%
and afterwords compute the negativity in accordance with (\ref{negativity}), 
\begin{equation}
\mathcal{N}_{jj^{\prime }}(t)=\left( \frac{2S+1}{4\pi }\right) ^{2}\int W_{%
\sqrt{\rho ^{t1}\rho ^{t1\dag }}}(\Omega _{j},\Omega _{j^{\prime
}}|t)d\Omega _{j}d\Omega _{j^{\prime }}-1  \label{N}
\end{equation}%
where $W_{\sqrt{\rho ^{t_{j}}\rho ^{t_{j}\dagger }}}(\Omega _{j},\Omega
_{j^{\prime }})=Tr(\sqrt{\rho _{jj^{\prime }}^{t_{j}}\rho _{jj^{\prime
}}^{t_{j}\dag }}\hat{\omega}(\Omega _{j})\hat{\omega}(\Omega _{j^{\prime }}))
$.

Curiosly, instead of the complicated expression for the negativity (\ref{N})
we have found that the following approximation can be used,\textbf{\ }%
\begin{equation}
\mathcal{N}(t)=\int d\Omega _{j}d\Omega _{j^{\prime }}|W^{tj}(\Omega
_{j},\Omega _{j^{\prime }}|t)|-1,  \label{NN}
\end{equation}%
where $W^{tj}(\Omega _{j},\Omega _{j^{\prime }})=Tr\left( \rho _{jj^{\prime
}}\hat{\omega}^{t}(\Omega _{j})\hat{\omega}(\Omega _{j^{\prime }})\right) $
is the Wigner symbol of the partially transposed density marix. We have
tested Eq.(\ref{NN}) for several randomly chosen density matrices.

One of the crucial ingredients of the phase-space approach is that the
Wigner function of the initial state can be efficiently approximated in the
limit $S\gg 1$ by using the asymptotic form of the mapping kernel \cite{wig
assym}. One can  show that the symbol of $j$-th spin state in  Eq.(\ref{psi0}%
),  $|\psi _{0}\rangle _{j}$, acquires the following approximated form\textbf{\ } 
\begin{equation}
\small{W_{j}(\Omega )\simeq \sum_{n,k=-S}^{S}\gamma _{k}\gamma
_{n}e^{-i\mu (k^{2}-n^{2})-i(n-k)\Phi +\frac{i\pi }{2}(n+k+2S)}d_{nk}^{S}(2%
\Theta )},  \label{Wapp}
\end{equation}%
where $d_{nk}^{S}(\Theta )$ is the Wigner $d$-function, and 
\begin{eqnarray*}
\mathrm{\cos }\Theta  &=&\mathrm{\cos }\theta \ \mathrm{\cos }\vartheta +%
\mathrm{\sin }\theta \sin \phi \sin \vartheta , \\
\mathrm{\sin }\Phi \mathrm{\sin }\Theta  &=&\mathrm{\sin }\theta \sin \phi
\cos \vartheta -\mathrm{\cos }\theta \ \mathrm{\sin }\vartheta ,
\end{eqnarray*}%
which essentially simplifies the simulation of the quantum evolution for
large spins.

\begin{figure}[tbp]
\centering
\includegraphics[width=7cm]{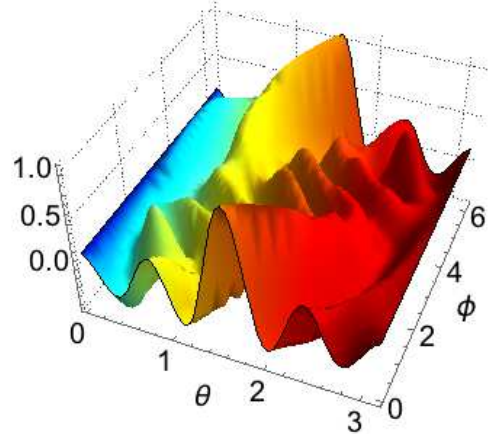}
\caption{The exact Wigner function of the initial spin state $|\psi _{0}\rangle _{j}$ in Eq.\eqref{psi0}%
.}
\label{fig:mil333}
\end{figure}
In Fig.(\ref{fig:mil333}) we plot the initial Wigner function, which has a
surprisingly non-trivial form (it was verified that the approximate
expression Eq. (\ref{Wapp}) describes well the exact one, Eqs. (\ref{Wex}), (%
\ref{eq:kernel})).

In the Figs. \ref{fig:mil31} - \ref{fig:mil32} we compare the quantum (solid
line) and the semi-classical (dotted line) dynamics. It is notable that a
very good agreement between both types of calculations even for long times
is observed. This allows to hope that the problem of optimization of the
correlation transfer in more complicated Ising and Heisenberg type large
spin chains can be appropriately analyzed by using the semi-classical
approach.

Finally, we have shown that an optimization procedure is required in order
to the improve the dynamic correlation transfer between different degrees of
freedom of large-spin chains. In order to perform such optimization the
phase-space methods can be quite useful, especially in the limit of
semiclassical-type systems.

\end{document}